\documentclass[11pt]{article}

\usepackage[a4paper,margin=29mm]{geometry}
\usepackage[T1]{fontenc}
\usepackage{lmodern}
\usepackage{microtype}
\usepackage{amsmath,amssymb,amsthm,mathtools}
\usepackage{enumitem}
\usepackage{xcolor}
\usepackage{listings}
\usepackage[hidelinks]{hyperref}

\setlist[itemize]{leftmargin=2em,itemsep=2pt,topsep=4pt}

\theoremstyle{definition}

\theoremstyle{remark}

\newcommand{\C}{\mathbb C}
\newcommand{\R}{\mathbb R}

\newcommand{\Lap}{\mathcal L}

\newcommand{\dd}{\,\mathrm d}
\newcommand{\e}{\mathrm e}
\newcommand{\lean}[1]{\texttt{#1}}
\DeclareMathOperator{\sinc}{sinc}

\definecolor{leanblue}{RGB}{36,72,122}
\definecolor{leangray}{RGB}{88,96,105}
\lstdefinelanguage{LeanFour}{
  morekeywords={def,theorem,lemma,noncomputable,structure,where,by,fun,let,have,
    exact,apply,rw,simpa,using,if,then,else,forall,section,end},
  sensitive=true,
  morecomment=[l]{--},
  morecomment=[s]{/-}{-/},
  morestring=[b]"
}
\title{A Formalization of the Laplace Transform and Its Inversion in Lean 4}

\author{
Daniel Goldberg\thanks{
  Department of Mathematics,
  Technion -- Israel Institute of Technology, Haifa, Israel.
  Email: \href{mailto:daniel.gold@campus.technion.ac.il}
  {\texttt{daniel.gold@campus.technion.ac.il}}
}
\and
Antoine Vinciguerra\thanks{
  Department of Computer Science,
  Technion -- Israel Institute of Technology, Haifa, Israel.
  Email: \href{mailto:antoine.v@campus.technion.ac.il}
  {\texttt{antoine.v@campus.technion.ac.il}}
}
}

\date{}

\begin{document}

\maketitle

\begin{abstract}
We present a Lean 4 formalization of the Laplace transform for
complex-valued functions, its fundamental operational rules, and a
Bromwich-type inversion theorem proved through real-variable integration and
the Dirichlet integral.  As an application, we formalize the Laplace-domain
solution of the harmonic oscillator and identify its transform with that of
$\sin(\omega t)$.  We also discuss the principal analytic and formalization
challenges encountered in the development.
\end{abstract}

\medskip 


\section{Introduction}

The Laplace transform converts a function of a real variable into a function of
a complex variable.  It is given by
\begin{equation}\label{eq:laplace}
  (\Lap f)(s)=\int_0^\infty f(t)\e^{-st}\dd t,
  \qquad s\in\C,
\end{equation}
whenever the integral exists. Algebraic operations in the
transform variable encode analytic operations in the original
variable; standard analytic treatments include those of Widder and
Doetsch~\cite{Widder1941,Doetsch1974}.  In particular, differentiation becomes
multiplication by $s$ up to
initial values,
\begin{equation}\label{eq:deriv-intro}
  \Lap(f')(s)=s\Lap f(s)-f(0),
\end{equation}
and translation of the transform variable corresponds to multiplication by an
exponential.  These identities explain the central role of the transform in the
study of differential and integral equations; they are among the classical rules developed systematically in
\cite{Doetsch1974,DebnathBhatta2015}.

The transform emerged over a long period rather than from a single isolated
definition.  Its prehistory includes Euler's integral-transform methods and
Laplace's use of related integral expressions in probability, while the
nineteenth- and early-twentieth-century development passed through complex
analysis, Heaviside's operational calculus, and the rigorous transform theory
associated with Bromwich and Doetsch. Deakin gives a detailed historical
account of this development~\cite{Deakin1981,Deakin1982}; primary landmarks
include Laplace's \emph{Th\'eorie analytique des probabilit\'es}~\cite{Laplace1812}
and Heaviside's \emph{Electromagnetic Theory}~\cite{Heaviside1893}.

The inverse is provided by the inverse Laplace transform.  If the $\operatorname{Re}s=\gamma$ lies in a suitable region of
convergence, the Bromwich formula takes the form
\begin{equation}\label{eq:bromwich}
  f(t)=\frac{1}{2\pi i}\int_{\gamma-i\infty}^{\gamma+i\infty}
       \e^{st}(\Lap f)(s)\dd s.
\end{equation}
This vertical-line inversion is conventionally associated with Bromwich; see
the original dynamical-systems paper~\cite{Bromwich1917} and the classical
treatments in \cite{Widder1941,Doetsch1974}.
In a classical proof one may close a contour and invoke the residue theorem.
We do not follow that route: the contour-integration and residue-theorem
infrastructure needed for this argument is not presently developed in the Lean~4
library to the required extent.  We therefore do not use complex contour
integration directly.  Instead, writing $s=\gamma+ir$ turns the vertical-line
expression into an integral over $r\in\R$, after which the proof uses real
parameter integrals, Fubini's theorem, and the Dirichlet integral.  The
measure-theoretic tools used in this passage are standard forms of the
Fubini--Tonelli and dominated-convergence theorems~\cite{Folland1999}.
Formula~\eqref{eq:bromwich} is deceptively compact.  A formal proof must specify
the meaning of the unbounded integral, the hypotheses permitting an exchange of
integrals, and the precise mode of
convergence of the oscillatory kernel.

Laplace transforms appear throughout applied mathematics.  They turn linear
ordinary differential equations with prescribed initial data into algebraic
equations, convert convolution into multiplication, and encode the response of
linear time-invariant systems.  They are used in the analysis of electrical
circuits, feedback and control systems, diffusion and wave propagation,
renewal-type equations, probability distributions, and signal processing.
The common theme is that evolution in time becomes algebra in the complex
frequency variable, while poles and regions of convergence retain information
about growth and oscillation~\cite{DebnathBhatta2015}.

The harmonic oscillator is the simplest nontrivial illustration.  For
$\omega\in\R$, consider
\begin{equation}\label{eq:oscillator}
  y''(t)+\omega^2y(t)=0,\qquad y(0)=0,\qquad y'(0)=\omega.
\end{equation}
Taking Laplace transforms gives
\[
  \bigl(s^2+\omega^2\bigr)\Lap y(s)=\omega.
\]
On the right half-plane $\operatorname{Re}s>0$ the denominator does not vanish,
and the resulting expression is the transform of $\sin(\omega t)$.  We treat
this calculation in detail, including the analytic hypotheses which are usually
suppressed in a textbook derivation.

The development is written in Lean~4~\cite{deMouraUllrich2021} and builds on the
measure-theoretic and analytic infrastructure of mathlib~\cite{Mathlib2020}.The complete formal development is available in
the accompanying repository \cite{laplacegithub}.

\subsection{Main contributions}

The principal contributions of the development are the following.

\begin{itemize}
\item We formalize a generalized Laplace kernel and its integral transform on a
  complete normed algebra, together with scalar and additive laws, and specialize
  this construction to the transform of complex-valued functions on
  the real line.
\item We formalize the usual transform as a Bochner integral over the
  positive half-line.  We prove convergence of truncated transforms and establish
  formulas for constants, powers, exponentials, derivatives, iterated
  derivatives, sine, cosine, exponential multiplication, and positive scaling,
  under explicit integrability and decay assumptions.
\item We define a Bromwich-type inverse transform and prove that it recovers a
  differentiable function at every point of a prescribed positive set.  The
  proof uses the formalisation of the Dirichlet integral developed in~\cite{GoldbergVinciguerraDirichlet}.
\item We give a formal Laplace-transform calculation for the harmonic oscillator.
  The theorem identifies the transform of any function satisfying the equation,
  initial conditions, regularity, integrability, and decay assumptions with the
  transform of $\sin(\omega t)$.
\end{itemize}

The formalization separates two useful concrete interfaces.  The first is the
specialization of the generalized transform through a measure pushed forward to
the real axis inside $\C$; it is used in the inversion theorem.  The second is a
direct set integral over $(0,\infty)$; it is convenient for truncated integrals,
operational rules, and the oscillator example.  Mathematically the endpoint at
zero is null, but keeping both presentations makes their respective proof roles
transparent.

The paper is organised as follows:
Section~\ref{sec:outline} gives a concise account of the underlying proofs, with
more detail for inversion and the oscillator.  Section~\ref{sec:formalization}
describes the Lean definitions and the structure of the development.
Section~\ref{sec:difficulties} discusses the parts of the formalization that
required the most care.  We conclude with the present scope of the library and
directions for extending it.

\section{Outline of the mathematical proofs}\label{sec:outline}

\subsection{The transform and its basic rules}

For a complex-valued function $f$ on $\R$, we regard the transform as
the Lebesgue integral of $f(t)\e^{-st}$ restricted to the positive half-line.
It is often useful to begin with the truncated transform
\begin{equation}\label{eq:finite-laplace}
  (\Lap_Tf)(s)=\int_{(0,T]} f(t)\e^{-st}\dd t.
\end{equation}
If $f(t)\e^{-st}$ is integrable on $(0,\infty)$, then
$\Lap_Tf(s)\to\Lap f(s)$ as $T\to+\infty$.  Indeed,
\[
  \Lap_Tf(s)=\int_{(0,\infty)}
     \mathbf 1_{(-\infty,T]}(t)f(t)\e^{-st}\dd t,
\]
and dominated convergence applies with as bound
$|f(t)\e^{-st}|$; we compare the scalar theorem in \cite{Folland1999} and its
Bochner-integral form in \cite{HytonenEtAl2016}.

Linearity follows directly from linearity of the Bochner integral
\cite{HytonenEtAl2016}.  Closed forms
for standard functions are first proved at finite $T$ and then passed to the
limit, following the usual transform-table derivations
\cite{DebnathBhatta2015,Doetsch1974}.  For example, when
$\operatorname{Re}s>0$,
\begin{align}
  \Lap 1(s)&=\frac1s,\label{eq:one}\\
  \Lap\!\left(\frac{t^k}{k!}\right)(s)&=\frac1{s^{k+1}},\label{eq:powers}\\
  \Lap(\e^{at})(s)&=\frac1{s-a}
       \quad\text{if }\operatorname{Re}a<\operatorname{Re}s.\label{eq:exponential}
\end{align}

The exponential representation of trigonometric functions gives
\begin{equation}\label{eq:sin-cos}
  \Lap(\cos(\omega t))(s)=\frac{s}{s^2+\omega^2},\qquad
  \Lap(\sin(\omega t))(s)=\frac{\omega}{s^2+\omega^2},
\end{equation}
for $\operatorname{Re}s>0$, again with the relevant integrability assumptions in
the formal statements.  The denominator is shown to be nonzero by the
factorization
\begin{equation}\label{eq:factor}
  s^2+\omega^2=(s-i\omega)(s+i\omega).
\end{equation}
Neither factor can vanish when $\operatorname{Re}s>0$.

\subsection{Derivatives}

On a bounded interval, integration by parts gives the standard derivative
rule~\cite{Doetsch1974}
\begin{equation}\label{eq:finite-deriv}
  \Lap_T(f')(s)
    =s\Lap_Tf(s)-f(0)+\e^{-sT}f(T).
\end{equation}
Provided the two Laplace integrands are integrable and the final boundary term
tends to zero, the limit $T\to\infty$ proves
Equation~\eqref{eq:deriv-intro}.  Iterating the same calculation yields
\begin{equation}\label{eq:nth-deriv}
  \Lap(f^{(n)})(s)=s^n\Lap f(s)
   -\sum_{k=0}^{n-1}s^{n-1-k}f^{(k)}(0).
\end{equation}
The finite identity contains an additional endpoint contribution
\[
  \e^{-sT}\sum_{k=0}^{n-1}s^{n-1-k}f^{(k)}(T),
\]
which is required to converge to zero.  Stating this hypothesis explicitly is
important in a formal setting: smoothness alone does not control growth at
infinity.  Classical transform texts impose comparable exponential-order or
boundary hypotheses before applying the rule~\cite{Widder1941,Doetsch1974}.

\subsection{The inverse transform}\label{subsec:inverse-outline}

The choice of proof is dictated in part by the available formal library.  The
usual complex-analytic derivation based on contour deformation and residues would
first require a substantial development of the relevant complex integration
theory.  The present proof avoids that detour.  Its only use of the vertical
complex line is the explicit parameterization $s=\gamma+ir$; from that point on,
the argument is formulated using integrals over real variables.  This leads
naturally to the oscillatory kernel and ultimately to the Dirichlet integral.
Both complex-analytic and Fourier-type approaches to Laplace inversion are
classical; see \cite{Bromwich1917,Widder1941,Doetsch1974}.

Fix $\gamma\in\R$.  Parameterizing the vertical line in
Equation~\eqref{eq:bromwich} by $s=\gamma+ir$ motivates the truncated inverse
\begin{equation}\label{eq:bounded-inverse}
  B_T(F)(t)=\frac{1}{2\pi}
     \int_{-T}^{T}\e^{(\gamma+ir)t}F(\gamma+ir)\dd r.
\end{equation}
The Lean definition retains the contour factor in the equivalent form
\[
  \frac{1}{2\pi i}\int_{-T}^{T}
    i\e^{(\gamma+ir)t}F(\gamma+ir)\dd r.
\]
Assume now that $F=\Lap f$.  Expanding the transform and exchanging the bounded
$r$-integral with the $a$-integral gives
\begin{align}
 B_T(\Lap f)(t)
 &=\frac{1}{2\pi}\int_0^\infty
     f(a)\e^{-\gamma(a-t)}
     \left(\int_{-T}^{T}\e^{ir(t-a)}\dd r\right)\dd a.\label{eq:fubini-result}
\end{align}
The exchange is justified by an explicit integrability proof on the product of
the restricted measure on $[-T,T]$ and the measure on the nonnegative real axis.
The compactness of $[-T,T]$ bounds the factor depending on $r$, while the assumed
integrability of $f(a)\e^{-\gamma a}$ controls the remaining factor.  This is
the standard quantitative hypothesis behind Fubini's theorem
\cite{Folland1999}.

For $a\ne t$ the inner integral is
\begin{equation}\label{eq:complex-dirichlet}
  \int_{-T}^{T}\e^{ir(t-a)}\dd r
      =\frac{2\sin(T(t-a))}{t-a}.
\end{equation}
For $a=t$ it is $2T$.  The singleton $\{t\}$ has Lebesgue measure zero, so
changing the integrand at that point does not change the outer integral.  With
the continuous convention $\sinc x=\sin x/x$ for $x\ne0$ and $\sinc 0=1$, one
obtains, by the usual almost-everywhere invariance of the Lebesgue integral
\cite{Folland1999},
\begin{equation}\label{eq:sinc-kernel}
  B_T(\Lap f)(t)=
  \int_0^\infty f(a)\e^{-\gamma(a-t)}
      \frac{T}{\pi}\sinc\bigl(T(t-a)\bigr)\dd a.
\end{equation}

The remaining limit is expressed through the normalized primitive of the sinc
function
\begin{equation}\label{eq:dirichlet-cutoff}
  D(x)=\frac12+\frac1\pi\int_0^x\sinc u\dd u.
\end{equation}
Our earlier formalization proves that
\begin{equation}\label{eq:heaviside-limit}
  D(Rx)\longrightarrow
  H(x):=
  \begin{cases}
    1,&x>0,\\
    \tfrac12,&x=0,\\
    0,&x<0,
  \end{cases}
  \qquad R\to+\infty,
\end{equation}
as a consequence of the Dirichlet integral
$\lim_{R\to\infty}\int_0^R\sinc u\dd u=\pi/2$; see
Dirichlet's foundational Fourier-series memoir~\cite{Dirichlet1829} and our
formal development~\cite{GoldbergVinciguerraDirichlet}.  In addition, $D$ is
bounded on $\R$.

Set
\[
  u(a)=f(a)\e^{-\gamma(a-t)}.
\]
Since $D'(T(a-t))=T\sinc(T(a-t))/\pi$, integration by parts rewrites
Equation~\eqref{eq:sinc-kernel} as
\begin{equation}\label{eq:inverse-ibp}
 B_T(\Lap f)(t)
 =-f(0)\e^{\gamma t}D(-Tt)
  -\int_0^\infty u'(a)D(T(a-t))\dd a.
\end{equation}
For $t>0$, the first term tends to zero by
Equation~\eqref{eq:heaviside-limit}.  The boundedness of $D$, the integrability
of $u'$, and dominated convergence give
\begin{align*}
 \int_0^\infty u'(a)D(T(a-t))\dd a
   &\longrightarrow \int_t^\infty u'(a)\dd a\\
   &=-u(t)=-f(t),
\end{align*}
where decay of $u$ at infinity follows from the integrability assumptions on
$u$ and $u'$; this is the one-dimensional absolutely-continuous/Sobolev
principle used at the unbounded endpoint~\cite{Brezis2011}.  The passage under
the integral is another application of dominated convergence~\cite{Folland1999}.
Thus $B_T(\Lap f)(t)\to f(t)$.

Finally, integrability of the inverse kernel implies independently that
$B_T(\Lap f)(t)$ tends to the unbounded inverse integral as
$T\to\infty$.  Uniqueness of limits identifies that integral with $f(t)$.  The
formal theorem is pointwise on a set $S\subset(0,\infty)$ and assumes continuity
and differentiability of $f$, integrability of $f(t)\e^{-\gamma t}$ and
$f'(t)\e^{-\gamma t}$, measurability of $f$, and integrability of the inverse
kernel for every $t\in S$.  These hypotheses are stronger and more explicit than
many classical presentations, but they align directly with the lemmas used in
the proof; compare the hypotheses in the classical inversion theories of
\cite{Widder1941,Doetsch1974}.

\subsection{The harmonic oscillator}\label{subsec:oscillator-outline}

Let $y:\R\to\C$ satisfy Equation~\eqref{eq:oscillator}, together with the
regularity, integrability, and boundary hypotheses needed for
Equation~\eqref{eq:nth-deriv} with $n=2$.  Write $Y(s)=\Lap y(s)$.  The proof
proceeds as follows.

First, the iterated-derivative formula yields
\[
 \Lap(y'')(s)=s^2Y(s)-sy(0)-y'(0)=s^2Y(s)-\omega.
\]
Second, the differential equation holds on $(0,\infty)$, hence the Laplace
transform of $y''+\omega^2y$ is zero.  Linearity and scalar multiplication then
give
\[
 0=s^2Y(s)-\omega+\omega^2Y(s),
\]
or equivalently
\begin{equation}\label{eq:oscillator-algebra}
 (s^2+\omega^2)Y(s)=\omega.
\end{equation}
Third, Equation~\eqref{eq:factor} and $\operatorname{Re}s>0$ show that the
coefficient in Equation~\eqref{eq:oscillator-algebra} is nonzero.  Therefore
\[
 Y(s)=\frac{\omega}{s^2+\omega^2}.
\]
The already established sine formula identifies the right-hand side with
$\Lap(\sin(\omega t))(s)$.  This is the standard operational solution of the
harmonic oscillator~\cite{DebnathBhatta2015}, with all analytic
side conditions exposed.

The conclusion formalized here is equality of the two Laplace transforms at the
chosen $s$. 

\section{Structure of the Lean formalization}\label{sec:formalization}

The four principal source files contain approximately $2{,}900$ lines of Lean,
in addition to the imported Dirichlet-integral module.  They are organized by
increasing specificity.
\begin{itemize}
\item \lean{LaplaceTransformDef.lean}: the abstract kernel and transform;
\item \lean{LaplaceTransformProperties.lean}: the direct transform
  and its operational rules;
\item \lean{RealLaplaceTransform.lean}: the specialization of the abstract
  construction and the inverse theorem;
\item \lean{LaplaceTransformExample.lean}: the harmonic-oscillator application.
\end{itemize}

\subsection{An abstract kernel}

The general construction is carried out in a complete normed algebra $E$ over
$\C$, using the standard Bochner-integral framework for Banach-space-valued
functions~\cite{HytonenEtAl2016}.  Given a set $S\subseteq E$, a map
$L:S\to\C\to E$, a function
$f:S\to E$, and a measure $\mu$ on $S$, the development defines the kernel
$\exp(-L(e,p))$, multiplies it by $f(e)$, and integrates:

\begin{lstlisting}
def laplaceKernel (S : Set E) (L : S -> C -> E)
    (e : S) (p : C) : E :=
  NormedSpace.exp C (-(L e p))

def fullLaplaceKernel (S : Set E) (L : S -> C -> E)
    (f : S -> E) (p : C) : S -> E :=
  fun e => f e * (laplaceKernel S L e p) * 1

def GeneralizedLaplaceTransform
    (S : Set E) (L : S -> C -> E) (f : S -> E)
    (mu : Measure S) : C -> E :=
  fun p => integral (fun e => fullLaplaceKernel S L f p e) mu
\end{lstlisting}

Here and below, code displays use ASCII transliterations for a few Lean symbols.
The actual source uses scalar multiplication in the kernel expression so that
the definition typechecks in the stated normed-algebra setting.  The main
results at this level are scalar linearity and additivity, each under the
integrability assumptions required by the Bochner integral.  The development
also proves a parameter-addition rule when $L(e,p_1+p_2)$ splits as
$L(e,p_1)+L(e,p_2)$ and the relevant elements commute.  This abstract layer is
small, but it isolates the algebra of the exponential kernel from the
measure-theoretic choices made later.

\subsection{The real axis and the measure}

For the specialization, the real line is represented as the subtype
\[
  \{z:\C\mid \operatorname{Im}z=0\}.
\]
The maps \lean{real\_to\_realLine} and
\lean{realLine\_to\_real} move between this subtype and $\R$, and the exponent
map is $L(x,z)=xz$.  Lebesgue measure restricted to $[0,\infty)$ is pushed
forward along the embedding $\R\to\C$.  The resulting definition is:

\begin{lstlisting}
def mu_real : Measure R := volume.restrict (Ici 0)
def mu_r : Measure realLine :=
  Measure.map real_to_realLine mu_real
\end{lstlisting}

\noindent The specialized transform is then defined by
\begin{lstlisting}
def RealLaplaceTransform (f : R -> C) : C -> C :=
  let g (x : realLine) : C := f (realLine_to_real x)
  GeneralizedLaplaceTransform realLine L g mu_r
\end{lstlisting}

The theorem \lean{RealLaplaceTransformIs} unfolds this construction into the
familiar integral
\[
  \operatorname{RealLaplaceTransform}(f)(p)
   =\int \e^{-pt}f(t)\dd\mu_{\!\R}(t).
\]
Its proof is longer than the mathematical identity suggests.  One must prove
measurability of the subtype embedding, measurability of the integrand on the
complex real axis, compatibility with the mapped measure, and a chain of
equalities removing coercions and commuting complex factors.  This theorem is
the bridge used when the inverse transform is expanded inside a double integral;
the underlying mapped-measure and change-of-variables facts are standard
measure theory~\cite{Folland1999}.

For operational rules, a second definition is used directly:

\begin{lstlisting}
noncomputable def laplaceTransform
    (f : R -> C) (s : C) : C :=
  integral (fun t => f t * cexp (-s * t))
    (volume.restrict (Ioi 0))

noncomputable def finiteLaplaceTransform
    (f : R -> C) (s : C) (T : R) : C :=
  integral (fun t => f t * cexp (-s * t))
    (volume.restrict (Ioc 0 T))
\end{lstlisting}

The two concrete formulations reflect two proof needs.  The mapped-measure
version is literally an instance of the general transform, whereas the direct
version works smoothly with interval integrals and existing improper-integral
lemmas.  The difference between $[0,\infty)$ and $(0,\infty)$ has measure zero;
nevertheless, Lean requires the relevant set-integral conversion whenever one
crosses from one presentation to the other.

\subsection{From bounded intervals to the full transform}

The theorem \lean{finite\_laplace\_tendsto\_laplace} is the analytic hinge for
many transform formulas.  It rewrites $(0,T]$ as
$(0,\infty)\cap(-\infty,T]$ and then expresses the truncated integrand by an
indicator function.  The proof applies the filter form of dominated convergence:
the indicator preserves almost-everywhere strong measurability, its norm is
bounded by that of the full integrand, and for each positive $t$ the indicator
is eventually equal to one as $T\to\infty$.  This is the filter-level analogue
of the usual scalar and Bochner dominated-convergence arguments
\cite{Folland1999,HytonenEtAl2016}.

This organization avoids repeating an improper-integral argument for every
transform pair.  Each finite formula is an exact interval identity.  The
corresponding infinite formula combines it with a convergence theorem for the
left-hand side and a separate limit calculation for the explicit right-hand
side.  Equality follows from uniqueness of limits in $\C$.

The derivative theorem follows the same pattern.  The finite proof uses the
fundamental theorem of calculus and integration by parts.  The infinite theorem
does not conceal the endpoint at infinity; it takes the convergence
$\e^{-sT}f(T)\to0$ as a hypothesis.  The higher-derivative result packages all
initial values into a finite sum indexed by \lean{Finset.range n}.  At $n=2$,
the oscillator proof explicitly unfolds this sum, an operation which is trivial
on paper but essential for Lean to see the two initial conditions.

\subsection{Standard transform pairs}

The exponential formula is proved by combining exponentials and using the
strict inequality $\operatorname{Re}a<\operatorname{Re}s$ to establish decay.
The sine and cosine theorems use their complex exponential representations.
These are the classical elementary transform-pair proofs
\cite{Doetsch1974,DebnathBhatta2015}.
The final field simplifications depend on separate proofs that
$s-i\omega$, $s+i\omega$, and $s^2+\omega^2$ are nonzero.  This is a typical
formalization pattern: algebraic normalization becomes reliable only after all
denominators have been discharged explicitly.

Two further structural identities are particularly short once definitions are
unfolded:
\begin{align*}
 \Lap\bigl(\e^{at}f(t)\bigr)(s)&=\Lap f(s-a),\\
 \Lap\bigl(f(at)\bigr)(s)&=\frac1a\Lap f(s/a),\qquad a>0.
\end{align*}
The second identity uses a change-of-variables theorem for the integral over the
positive half-line and then reconciles real and complex inverses by coercion;
compare the standard scaling rule in \cite{Doetsch1974}.

\subsection{Encoding the inverse transform}

The vertical line is parameterized by

\begin{lstlisting}
def imNbFromReals (gamma T : R) : C := gamma + T * I

def InverseLaplaceKernel (F : C -> C) (t : R) : R -> R -> C :=
  fun gamma T => I * cexp (imNbFromReals gamma T * t)
                    * F (imNbFromReals gamma T)

def inverseLaplace_t (F : C -> C) (gamma t : R) : C :=
  1 / (2 * I * pi) * integral (InverseLaplaceKernel F t gamma)
\end{lstlisting}

There is also a bounded version integrating over $[-T,T]$.  At the function
level, the definitions take a set $S\subseteq\R$ and a proof that the inverse
kernel is integrable for every $t\in S$.  The vertical-line parameterization is
the real-variable form of the Bromwich contour
\cite{Bromwich1917,Doetsch1974}:

\begin{lstlisting}
def inverseLaplaceFunction (F : C -> C) (gamma : R) (S : Set R)
    (h : forall t in S, Integrable (InverseLaplaceKernelFunctT F t gamma))
    : S -> C :=
  fun t => inverseLaplace_t F gamma t
\end{lstlisting}

The proof argument records the natural domain of the partial analytic
construction while the returned function itself has a simple subtype domain.
Additivity and scalar multiplication are proved first for the kernel and then
for the inverse integral.  The theorem
\lean{limit\_inverseLaplace\_bounded\_eq\_full} shows that bounded symmetric
integrals converge to the full inverse by dominated convergence.

\subsection{The bounded inversion identity}

The central intermediate result is
\lean{IsInverseLaplaceBounded}.  In mathematical notation it states
Equation~\eqref{eq:sinc-kernel}.  Its Lean proof makes each informal interchange
and simplification into a separate lemma.

The lemma \lean{Fubini\_lemma} exchanges the $r$- and $a$-integrals.  Its
hypotheses include measurability of $f$, integrability of the exponentially
weighted function, and integrability of the two-variable integrand with respect
to the product measure, precisely as required by the Fubini--Tonelli theorem
\cite{Folland1999}.  Inside \lean{IsInverseLaplaceBounded}, the latter is
constructed rather than assumed: continuity yields a bound for the vertical-line
factor on the compact interval $[-T,T]$, and the weighted integrability of $f$
then gives an integrable product majorant.

The lemma \lean{integral\_cexp\_Icc\_Dirichlet} proves the case distinction in
Equation~\eqref{eq:complex-dirichlet}.  The zero-frequency case reduces to the
measure of $[-T,T]$; the nonzero case uses an antiderivative for the complex
exponential and simplifies the numerator to a sine.  The follow-up lemma
\lean{h2ndIntegralCalc} removes the exceptional point $a=t$ from the outer
integral by observing that the singleton $\{t\}$ has measure zero.  Finally,
\lean{integral\_sinc\_equivalence} converts the quotient to the continuous sinc
kernel almost everywhere.

\subsection{From the Dirichlet integral to inversion}

The next theorem, \lean{IsInverseLaplaceBounded'}, is the formal integration by
parts identity~\eqref{eq:inverse-ibp}.  The derivative
is established separately:
\[
 \frac{\dd}{\dd a}D(T(a-t))
   =\frac{T}{\pi}\sinc(T(t-a)),
\]
where evenness of sinc reconciles the sign.  The derivative of
$f(a)\e^{-\gamma(a-t)}$ is also isolated as a lemma.  The proof then needs
integrability of both products appearing in integration by parts, continuity at
the finite endpoint, and vanishing of the weighted function at the infinite
endpoint.  The analytic pattern is the absolutely-continuous integration by
parts used in one-dimensional Sobolev theory~\cite{Brezis2011}.

The vanishing statements are not inferred from integrability of the function
alone.  The development proves them from differentiability together with
integrability of the weighted function and of its derivative.  This is one of
the important places where the formal proof exposes a hypothesis often hidden in
an informal phrase such as ``the boundary term vanishes.''

The convergence theorem
\lean{Tendsto\_Dirichlet\_Integral} applies the previously formalized cutoff
limit, ultimately based on the Dirichlet integral
\cite{Dirichlet1829,GoldbergVinciguerraDirichlet}, to the derivative of the
weighted function and obtains the limit $-f(t)$.
The separate lemma
\lean{Tendsto\_Dirichlet\_Integral\_times\_const} sends the endpoint term at
zero to zero when $t>0$.  The final theorem can therefore be summarized by the
following ASCII rendering of its statement:

\begin{lstlisting}
theorem IsInverseLaplace
    (f : R -> C) (gamma : R) (S : Set R)
    (hS : forall x in S, 0 < x)
    (h_cont : Continuous f)
    (h_int : Integrable (fun t => f t * cexp (-(gamma * t))))
    (h_meas : Measurable f)
    (h_inv : forall t in S,
      Integrable (InverseLaplaceKernelFunctT (RealLaplaceTransform f) t gamma))
    (h_diff : Differentiable R f)
    (h_diff_int : Integrable (fun t => deriv f t * cexp (-gamma * t))) :
    forall t : S,
      inverseLaplaceFunction (RealLaplaceTransform f) gamma S h_inv t = f t
\end{lstlisting}

The proof constructs two limits of the same bounded inverse.  One limit is the
full inverse integral; the other is $f(t)$.  The result follows from
\lean{tendsto\_nhds\_unique}.  This final use of uniqueness neatly isolates the
two analytic halves of the argument.

\subsection{The harmonic-oscillator theorem}

The application file states the equation on the open positive half-line.  Its
initial data are represented by the zeroth and first iterated derivatives at zero.
Its assumptions mirror exactly the inputs of the general transform lemmas:
$C^2$ regularity, integrability of $y$ and $y''$ after multiplication by the
Laplace kernel, convergence of the order-two endpoint sum, and integrability of
the two exponential modes used to evaluate sine.

The proof has seven explicit stages.  It invokes the iterated-derivative formula,
unfolds the two-term initial-value sum, proves that the transform of the
differential equation is zero almost everywhere on $(0,\infty)$, applies
linearity, moves $\omega^2$ outside the integral, solves the resulting algebraic
equation after proving the denominator nonzero, and finally rewrites with the
sine-transform theorem.  The conclusion is represented by:

\begin{lstlisting}
theorem harmonic_oscillator_laplace_eq ... :
  laplaceTransform y s =
  laplaceTransform
    (fun t => (cexp (I * omega * t) - cexp (-I * omega * t))
              / (2 * I)) s := by
  -- L(y'') = s^2 L(y) - omega
  -- L(y'' + omega^2 y) = 0
  -- (s^2 + omega^2) L(y) = omega
  -- compare with the proved transform of sine
  ...
\end{lstlisting}

The use of the exponential representation, rather than a separate real sine
function, matches the complex codomain of the transform and permits direct reuse
of the exponential transform rule.

\section{Difficult parts and design lessons}\label{sec:difficulties}

\subsection{Improper integrals are limits, not notation}

The first recurring difficulty is the boundary between Lebesgue integration and
improper integration.  Mathlib's integral is total as a function of the
integrand, so an expression can be syntactically meaningful even when the desired
integrability hypothesis is absent.  For that reason, every theorem records the
integrability needed to make its analytic interpretation valid.  The distinction
between Lebesgue integrals and limits of truncated integrals is standard in
measure theory~\cite{Folland1999}; the implementation rests on mathlib's
Bochner-integral infrastructure~\cite{Mathlib2020,HytonenEtAl2016}.

The transform itself is a set integral over the positive half-line.  Truncation
is not merely a change of notation: the proof must express a moving domain by an
indicator function and verify dominated convergence.  Similarly, the inverse
integral over the whole real line is reached through symmetric bounded
intervals.  These choices make the limiting behavior explicit and prevent a
conditionally convergent expression from being mistaken for a Lebesgue integral
\cite{Folland1999}.

The Dirichlet integral is a particularly relevant example.  The sinc function is
not absolutely integrable on the positive half-line, so its integral cannot be
used as an ordinary unbounded Lebesgue integral.  The result imported by the
Laplace development is instead a theorem about the limit of bounded interval
integrals.  That formulation, established in
\cite{Dirichlet1829,GoldbergVinciguerraDirichlet}, is exactly what the inversion
proof needs to obtain the pointwise cutoff limit.

\subsection{Fubini requires a quantitative proof}

Informal derivations of the Bromwich formula often exchange the Laplace integral
and the contour integral in one line.  In Lean, the two-variable function, the
product measure, and its integrability must be given explicitly.  The relevant
integrand is
\[
 (r,a)\longmapsto
 i\e^{(\gamma+ir)t}\e^{-(\gamma+ir)a}f(a).
\]
On $r\in[-T,T]$, the norm of the first exponential factor is bounded by a
constant depending on $t$ and $\gamma$ but not on $a$.  The oscillatory factors
have norm one, leaving an $a$-majorant proportional to
$|f(a)\e^{-\gamma a}|$.  The compact interval has finite measure, so the product
majorant is integrable.  Formalizing this argument required continuity and
compactness lemmas, construction of the bound, measure restriction estimates,
and the product-integrability API before the integral swap itself became a
single application of Fubini~\cite{Folland1999}.

This bounded-first approach is decisive.  Attempting to exchange both unbounded
integrals at once would demand substantially stronger absolute integrability and
would obscure the oscillatory nature of inversion.  Truncation provides precisely
the compactness needed for the estimate.

\subsection{Endpoint terms and decay}

Both the forward and inverse developments rely on integration by parts at an
unbounded endpoint.  In each case, the formal proof separates the finite identity
from the limiting boundary statement.  For the forward transform the explicit
hypothesis is an exponentially weighted endpoint sum.  For inversion, decay of
$u(a)=f(a)\e^{-\gamma(a-t)}$ at infinity is derived from integrability of $u$ and
$u'$.  The necessary results are stated for filters \lean{atTop} and
\lean{atBot}; constant exponential factors introduced by shifting $a-t$ are
handled by limit-preserving multiplication.  Analytically, this is the
one-dimensional absolutely-continuous decay principle discussed in
\cite{Brezis2011}.

The proof of inversion also has a genuine finite boundary at zero.  It survives
integration by parts as $-f(0)\e^{\gamma t}D(-Tt)$ and only disappears in the
limit because $t>0$.  This explains why the final theorem uses a subtype
$S\subset(0,\infty)$. 

\subsection{Coercions, measures, and normal forms}

The development moves frequently among real numbers, complex numbers, the
subtype of complex numbers with zero imaginary part, and functions on those
types.  Identities such as
\[
  -(\gamma a)=(-\gamma)a,
  \qquad \e^{x+y}=\e^x\e^y,
  \qquad (a^{-1}:\R)=(a:\C)^{-1}
\]
are mathematically elementary, but they do not always share the same syntactic
normal form after coercions.  Much of the local proof work consists of choosing a
normal form before invoking a library theorem or an algebraic tactic.

Mapped measures introduce an additional layer.  To turn the generalized
transform into the conventional formula, the proof establishes measurability of
both directions between $\R$ and the real-axis subtype and then applies the
integral formula for a mapped measure.  The result justifies the abstract
architecture, but it also suggests that a future refactoring could expose a
single canonical transform and prove the other presentation equal to
it. Mapped measures and the corresponding integration formula
are treated in \cite{Folland1999}.

\subsection{Algebraic automation after analytic preparation}

The final oscillator calculation is algebraically simple, but reliable
automation depends on prior analytic preparation.  The ODE is first converted
to an almost-everywhere zero integrand on the restricted measure.  Linearity is
applied only after the integrability of $\omega^2y$ has been derived.  Division
is performed only after nonvanishing of $s^2+\omega^2$ has been proved from
$\operatorname{Re}s>0$.  Once these obligations are discharged, ring and field
normalization close the remaining equalities.

\section{Scope of the present manuscript}

The manuscript concerns the definitions and proved results used for the forward
transform, the inverse theorem, and the oscillator but does not cover the properties of the Laplace Transform in tis full generality.\\
Convolution is a natural next operational rule, but it requires its own
careful treatment.

\section*{Acknowledgements}

The authors would like to thank Professor Yuval Filmus for being at the origin
of this project and for encouraging its development. We are also grateful to
Ashvni Narayanan for her valuable advice.

\end{document}